\documentclass[10pt, conference, compsocconf]{IEEEtran}

\usepackage{amsmath}
\usepackage{color}
\usepackage[table]{xcolor}
\usepackage{paralist}
\usepackage{comment}
\usepackage{listings}
\usepackage{xspace} 
\usepackage{url}
\usepackage{multirow}
\usepackage[bookmarks=false]{hyperref}
\usepackage[capitalize]{cleveref}
\usepackage{graphicx}
\usepackage[caption=false,font=footnotesize]{subfig}
\usepackage{afterpage}
\usepackage{tikz}

\newcommand{\tsdesc}[1]{\textcolor{red}{\textbf{#1}}}
\newcommand{\eg}{e.g.}
\newcommand{\ie}{i.e.}
\newcommand{\cf}{\textit{cf.}\xspace}

\newcommand{\pts}{\textit{pts}\xspace}
\newcommand{\wts}{\textit{wts}\xspace}
\newcommand{\rts}{\textit{rts}\xspace}
\newcommand{\dwts}{\textit{delta\_wts}\xspace}
\newcommand{\drts}{\textit{delta\_rts}\xspace}
\newcommand{\dts}{\textit{delta\_ts}\xspace}
\newcommand{\mts}{\textit{mts}\xspace}

\newcommand{\bts}{\textit{bts}\xspace}

\newcommand{\data}{D\xspace}
\newcommand{\msg}{M\xspace}
\newcommand{\notimplies}{%
  \mathrel{{\ooalign{\hidewidth$\not\phantom{=}$\hidewidth\cr$\implies$}}}}\newcommand{\tsman}{timestamp 
  manager\xspace}

\newtheorem{definition}{Definition}

\makeatletter
\def\@copyrightspace{\relax}
\makeatother

\begin{document}


\title{Tardis: Time Traveling Coherence Algorithm for Distributed Shared Memory}

\author{
\IEEEauthorblockN{Xiangyao Yu}
\IEEEauthorblockA{CSAIL, MIT\\
Cambridge, MA, USA\\
yxy@mit.edu}
\and
\IEEEauthorblockN{Srinivas Devadas}
\IEEEauthorblockA{CSAIL, MIT\\
Cambridge, MA, USA\\
devadas@mit.edu}
}

\maketitle

\begin{abstract}

A new memory coherence protocol, Tardis, is proposed.
Tardis uses timestamp counters representing logical
time as well as physical time to order memory operations and enforce
sequential consistency in any type of shared memory system. Tardis
is unique in that as compared to the widely-adopted directory coherence protocol,
and its variants, it completely avoids multicasting and only 
requires $O(\log N)$ storage per cache block for an $N$-core system rather than
$O(N)$ sharer information. Tardis is simpler and easier to reason about, yet
achieves similar performance to directory protocols on a wide range of benchmarks run
on 16, 64 and 256 cores.


\end{abstract}


\section{Introduction} \label{sec:intro}

Shared memory systems are ubiquitous in parallel computing.
Examples include multi-core and multi-socket processors, and 
distributed shared memory systems (DSM). The correctness of these 
systems is defined by the memory consistency model which specifies the 
legitimate interleaving of operations from different nodes (\eg, cores 
or processors). The enforcement of a consistency model heavily relies 
on the underlying coherence protocol.  For a shared memory system, the 
coherence protocol is the key component to ensure performance and 
scalability.

When the data can be cached in the local memory of a node, most large-scale
shared memory systems today adopt directory based
coherence protocols~\cite{censier1978, tang1976}. Examples include 
many-core systems with large core count~\cite{tilera, xeonphi},
coherence between multi-socket systems like Intel's 
QPI~\cite{ziakas2010} and AMD's HyperTransport~\cite{anderson2003}, 
and coherence of distributed shared memory systems like 
IVY~\cite{li1989} and Treadmarks~\cite{keleher1994}.
A well known challenge in a directory coherence protocol is latency and 
scalability.  For example, these protocols 
keep a list of nodes (sharers) caching each data and send 
invalidations to sharers before the data is modified by some node.
Waiting for all invalidation requests to be acknowledged 
may take a long time and storing the sharer information or supporting 
broadcasting does not scale well as the number of nodes increases. 


We propose a new coherence protocol, Tardis, which is simpler and more 
scalable than the simplest directory protocol, but has equivalent performance.
Tardis {\em directly} expresses the memory consistency model by explicitly 
enforcing the global memory order using timestamp counters that represent
logical as opposed to physical time; it does this
without requiring a globally synchronized clock unlike prior timestamp 
coherence schemes (e.g., \cite{lis2011, singh2013}), and without
requiring multicast/broadcast support unlike prior directory
coherence schemes (e.g., \cite{ATAC, sanchez2012}).
In Tardis, only the timestamps and the owner ID need to be stored for 
each address for a $O(\log N)$ cost where $N$ is the number of 
processors or cores; the $O(N)$ sharer information of common
directory protocols is not required.
The requirement of storing sharers is avoided partly through the novel insight that
a writer can instantly jump ahead\footnote{hence the name Tardis!} to a time when the sharer copies have expired
and immediately perform the write without violating sequential consistency.
A formal proof that Tardis satisfies sequential consistency can be found in \cite{tardis-proof}.\footnote{The proof corresponds to a slightly simplified version of the protocol presented here.}

We evaluated Tardis in the context of multi-core processors.  Our 
experiments showed that Tardis achieves similar performance 
to its directory counterpart over a wide range of benchmarks. Due to 
its simplicity and excellent performance, we believe Tardis is a 
competitive alternative to directory coherence for 
massive-core and DSM systems.

We provide background in \cref{sec:background}, describe the basic 
Tardis protocol in \cref{sec:tardis}, and optimizations to the basic 
protocol in \cref{sec:opt}.
We evaluate Tardis in \cref{sec:eval}, 
discuss related work in \cref{sec:related} and conclude the paper in 
\cref{sec:conclusion}.

\section{Background} \label{sec:background}

In this section, we provide some background on memory consistency and 
coherence.

%
%
%

\subsection{Sequential Consistency} \label{sec:consistency}

A memory consistency model defines the correctness of a shared memory 
system. Specifically, it defines the legitimate behavior of memory 
loads and stores. Although a large number of consistency models exist, 
we will focus on {\it sequential consistency} due to its simplicity. 

Sequential consistency was first proposed and formalized by
Lamport~\cite{lamport1979}. A parallel program is sequentially 
consistent if
{\it ``the result of any execution is the same as if the operations of 
all processors (cores) were executed in some sequential order, and the 
operations of each individual processor (core) appear in this sequence 
in the order specified by its program''}. If we use $<_p$ and $<_m$ to 
denote program order and global memory order respectively, sequential 
consistency requires the following two rules to be 
held~\cite{weaver1994}: 

{\bf Rule 1:} $X <_{p} Y \implies X <_{m} Y $

{\bf Rule 2:}

\noindent${\it Value\ of}\ L(a) = {\it Value\ of\ Max}_{<m} \{S(a) | S(a) <_{m} 
L(a) \}$\\
where $L(a)$ is a load to address $a$ and $S(a)$ is a store to address 
$a$; the $Max_{<m}$ operator selects the most recent operation in the 
global memory order.

Rule 1 says that if an operation X (a load or a store) is before 
another operation Y in the program order of any core, X must precede Y 
in the global memory order. Rule 2 says that a load to an address 
should return the value of the most recent store to that address with 
respect to the global memory order. 

\subsection{Directory-Based Coherence} \label{sec:coherence}

In practical systems, each core/processor has some private local 
storage to exploit locality.
A memory coherence protocol is therefore used to support the 
consistency model. 

Although both snoopy and directory protocols are used in small 
systems, virtually all large-scale shared memory systems today use 
some variant of the basic directory-based coherence protocol. The directory is a 
software or hardware structure tracking how the data are shared or 
owned by different cores/processors. In a directory protocol, the 
second rule of sequential consistency is achieved through the 
invalidation mechanism; when a core/processor writes to an address 
that is shared, all the shared copies need to be invalidated before 
the write can happen.  Future reads to that address have to send 
requests to the directory which returns the value of the last write.  
This mechanism essentially guarantees that reads that happen after the 
last write with respect to physical time can only observe the value of 
the last write (the second rule of sequential consistency). 

The directory needs to keep the sharer information of each address in 
order to correctly deliver the invalidations. If the system has $N$ 
cores/processors, the canonical protocol requires $O(N)$ storage per 
address, which does not scale well when the system gets bigger.  
Alternative solutions to avoid $O(N)$ storage do exist (\cf 
\cref{sec:related}) but either require broadcasting, incur 
significant additional complexity, or do not perform well.


\section{Basic Protocol} \label{sec:tardis}

We present a new coherence protocol, Tardis, which only requires 
$O(\log N)$ storage per cacheline
and requires neither broadcasting/multicasting support nor a globally synchronized 
clock across the whole system. 
Tardis works for all types of distributed shared memory systems and is 
compatible with different memory consistency models. 

In this paper, we present the 
Tardis protocol for sequential consistency in the context of a multi-core processor with shared 
last level cache (LLC). Our discussion applies equally well to other 
types of shared memory systems. 


\subsection{Timestamp Ordering} \label{sec:tardis-tsorder}

In a directory protocol~(\cf \cref{sec:coherence}), the global memory 
order ($<_m$) is enforced through the physical time order. \ie, if $X$ 
and $Y$ are memory operations to the same address $A$ and one of them 
is a store, then 

\vspace{-0.1in}
\[\ X <_m Y ~~\implies~~ X <_{pt} Y \]
\vspace{-0.2in}

In Tardis, we break the correlation between the global memory order 
and the physical time order for \textit{write after read} (WAR) 
dependencies while maintaining the correlation for \textit{write after 
write} (WAW) and \textit{read after write} (RAW) dependencies. 

\vspace{-0.15in}
\[\ S_1(A) <_m S_2(A) \implies S_1(A) <_{pt} S_2(A) \]
\vspace{-0.2in}
\[\ S(A) <_m L(A) \implies S(A) <_{pt} L(A) \]
\vspace{-0.2in}
\[\ L(A) <_m S(A) \notimplies L(A) <_{pt} S(A) \]
\vspace{-0.2in}

Tardis achieves this by explicitly assigning a timestamp to each 
memory operation to indicate its global memory order.  Specifically, 
the global memory order in Tardis is defined as a combination of 
physical time and logical timestamp
order, i.e., physi-logical time order, which we will
call \textit{physiological time order} for ease of pronunciation.




%


\begin{definition}[Physiological Time Rule] \label{def:mem-order}
\vspace{-0.05in}
\[ X <_m Y ~~:=~~   X <_{ts} Y~or~(X =_{ts} Y~and~X <_{pt} Y) \]
\label{def0:mem-order}
\end{definition}
\vspace{-0.2in}

In Definition \ref{def0:mem-order} the global memory order is explicitly 
expressed using timestamps. Operations without dependency (e.g., 
two concurrent read operations) or with obvious relative ordering 
(e.g., accesses to private data from the same core) can share 
the same timestamp and their global memory order is implicitly 
expressed using the physical time order.

Using the physiological time rule, Rule 1 of sequential consistency 
becomes $X <_p Y \Rightarrow X <_{ts} \vee (X=_{ts} Y \wedge 
X<_{pt} Y)$. Assuming a processor always does in-order commit, we have $X 
<_p Y \Rightarrow X<_{pt}Y$. So Tardis only needs to guarantee that $X 
<_p Y \Rightarrow X\leq_{ts} Y$, i.e., operations from the same 
processor have monotonically increasing timestamps in the program order.  
For Rule 2 of sequential consistency, Tardis needs to guarantee
that a load observes the correct store in the global memory order as
defined by Definition \ref{def0:mem-order}. The correct store is the
latest store -- either the one with the largest logical timestamp or the
latest physical time among the stores with the largest logical timestamp~\cite{tardis-proof}.

We note that the physiological timestamp here is different from 
Lamport clocks \cite{lamport1978}. In Lamport clocks, a timestamp is 
incremented for each operation while a physiological timestamp is not 
incremented if the order is implicit in physical time. That said, 
the physiological timestamp does share some commonality with the Lamport 
clock. In a sense, Tardis applies Lamport/physiological timestamp to distributed 
shared memory systems.

\subsection{Tardis without Private Cache} \label{sec:tardis-noprmem}

In Tardis, timestamps are maintained as logical counters. Each core 
keeps a program timestamp (\pts) which is the timestamp of the last 
operation in the program order. Each cacheline keeps a read timestamp 
(\rts) and a write timestamp (\wts). The \rts equals the largest 
timestamp among all the loads of the cacheline thus far and the \wts 
equals the timestamp of the latest store to the cacheline. Tardis 
keeps the invariant that for a cacheline, its current data must be 
valid between its current \wts and \rts. The \pts should not be 
confused with the processor clock, it does not increment every cycle 
and is not globally synchronized.  The directory structure is replaced 
with a \tsman. Any load or store request to the LLC should go to the 
\tsman.  

For illustrative purposes, we first show the Tardis protocol assuming 
no private cache and all data fitting in the shared LLC. Each 
cacheline has a unique copy in the LLC which serves all the memory 
requests. Although no coherence protocol is required in such a system, 
the protocol in this section provides necessary background in 
understanding the more general Tardis protocol in 
\cref{sec:tardis-withprmem}. 

\begin{table}
	\caption{ Timestamp management in the Tardis Protocol without Private 
	Memory}
    \begin{center}
	{ \footnotesize
		\begin{tabular}{|p{0.7in}||p{1in}|p{1.2in}|}
			\hline
			Request Type & Load Request & Store Request \\ \hline
			Timestamp \newline Operation &
				   $\pts \Leftarrow Max(\pts, \wts)$\newline
				   $\rts \Leftarrow Max(\pts, \rts)$
				& $\pts \Leftarrow Max(\pts, \rts + 1)$\newline
				   $\wts \Leftarrow \pts $\newline
				   $\rts \Leftarrow \pts$ \\ \hline
		\end{tabular}
    }
    \end{center}
	\label{tab:noprmem}
	\vspace{-0.25in}
\end{table}

\cref{tab:noprmem} shows one possible timestamp management policy that 
obeys the two rules of sequential consistency. But other policies also 
exist. Each memory request contains the core's \pts before the current 
memory operation.  After the request, \pts is updated to the timestamp 
of the current operation.  

For a load request, the \tsman returns the value of the last store.
According to Rule 1, the load timestamp must be no less than the 
current \pts. According to Rule 2, the load timestamp must be no less 
than \wts which is the timestamp of the last store to this cacheline.  
So the timestamp of the load equals $Max(\pts, \wts)$.
If the final $\pts > \rts$, then \rts bumps up to this \pts since the \rts should be 
the timestamp of the last read in the timestamp order.

For a store request, the last load of the cacheline (at \rts) did not 
observe the value of the current store. According to Rule 2, the 
timestamp of the current store must be greater than the \rts of the 
cacheline (the timestamp of the last load). So \pts becomes $Max(\pts, 
\rts+1)$. \wts and \rts should also bump up to this final \pts since a 
new version has been created.

Both Rule 1 and Rule 2 hold throughout the protocol: the \pts 
never decreases and a load always observes the correct store in the 
timestamp order. 

\subsection{Tardis with Private Cache} \label{sec:tardis-withprmem}

With private caching, data accessed from the LLC are stored in the 
private cache. The protocol introduced in \cref{sec:tardis-noprmem} 
largely remains the same. However, two extra mechanisms need to be 
added.


{\bf Timestamp Reservation}: Unlike the previous protocol where a load 
happens at a particular timestamp, timestamp reservation allows a load 
to reserve the cacheline in the private cache for a period of logical 
time (\ie, the lease). The end timestamp of the reservation is stored 
in \rts.
The cacheline can be read until the timestamp expires (\pts $>$ \rts).  
If the cacheline being accessed has already expired, a request must be 
sent to the \tsman to extend the lease.

{\bf Exclusive Ownership}: Like in a directory protocol, a modified  
cacheline can be exclusively cached in a private cache.  In the 
\tsman, the cacheline is in exclusive state and the owner of the 
cacheline is also stored which requires $\log(N)$ bits of storage.
The data can be accessed freely by the owner core as long as it is in 
the exclusive state; and the timestamps are properly updated with each 
access.  If another core later accesses the same cacheline, a write 
back (the owner continues
to cache the line in shared state) or flush request (the owner 
invalidates the line) is sent to the owner which replies with the 
latest data and timestamps.

Note that in the private cache, the meanings of \rts for shared and 
exclusive cachelines are different. For a shared cacheline, \rts is 
the end timestamp of the reservation; for an exclusive cacheline, \rts 
is the timestamp of the last load or store.  The state transition and the 
timestamp management of Tardis with private cache are shown in 
\cref{tab:prmem} and \cref{tab:tm}.
\cref{tab:prmem} shows the state transition at the private cache and 
\cref{tab:tm} shows the state transition at the shared \tsman.  
\cref{tab:msg} shows the network message types used in the Tardis 
protocol where the suffix REQ and REP represent request and response 
respectively.

\begin{table*}[ht!]
	\caption{ State Transition in Private Cache. \textit{TM} is the
	shared 	\tsman, $D$ is the data, $M$ is the message, $reqM$ is the
	request message if two messages are involved. Timestamp transition
	is highlighted in \tsdesc{red}.
	}\begin{center}
	{ \scriptsize
		\begin{tabular}{|p{0.45in}||p{.85in}|p{.85in}|p{0.8in}||p{.7in}|p{.8in}|p{1.02in}|}
			\hline
			\multirow{2}{*}{States} & \multicolumn{3}{c||}{Core
			Event} & \multicolumn{3}{c|}{Network Event} \\ \cline{2-7}
	
			& Load & Store & Eviction & SH\_REP or \newline EX\_REP & RENEW\_REP or \newline UPGRADE\_REP & FLUSH\_REQ or \newline WB\_REQ \\ \hline \hline

			Invalid 
			  & send SH\_REQ to TM\newline
				\tsdesc{\msg.wts$\Leftarrow$0, \newline \msg.pts$\Leftarrow$pts}
			  & send EX\_REQ to TM\newline \tsdesc{\msg.wts$\Leftarrow$0} 
			  & \cellcolor[rgb]{.7,.7,.7} 
			  & \multirow{3}{0.9in}{Fill in data\newline 				
				\underline{\textbf{SH\_REP}} \newline
				\tsdesc{\data.wts$\Leftarrow$\msg.wts} \newline
				\tsdesc{\data.rts$\Leftarrow$\msg.rts} \newline
				state$\Leftarrow$Shared \newline
				\underline{\textbf{EX\_REP}} \newline				
				\tsdesc{\data.wts$\Leftarrow$\msg.wts} \newline
				\tsdesc{\data.rts$\Leftarrow$\msg.rts} \newline
				state$\Leftarrow$Excl.}
			  & \cellcolor[rgb]{.7,.7,.7} 
			  & \cellcolor[rgb]{.7,.7,.7}
			\\\cline{1-4} \cline{6-7}

			Shared \newline $\pts \leq \rts$
			  & Hit \newline \tsdesc{pts$\Leftarrow$Max(pts, \data.wts)}
			  & \multirow{2}{.9in}{send EX\_REQ to TM \newline
				\tsdesc{\msg.wts$\Leftarrow$\data.wts}}
			  & \multirow{2}{0.8in}{\cellcolor[rgb]{.8,.9,.8}state$\Leftarrow$Invalid \newline 
				No msg sent.}
			  & & \multirow{2}{.9in}{\underline{\textbf{RENEW\_REP}} \newline
				\tsdesc{\data.rts$\Leftarrow$\msg.rts} \newline
				\underline{\textbf{UPGRADE\_REP}} \newline
				\tsdesc{\data.rts$\Leftarrow$\msg.rts} \newline
				state$\Leftarrow$Excl.
				}
			  & \cellcolor[rgb]{.7,.7,.7}
			\\\cline{1-2} \cline{7-7}
			
			Shared \newline $\pts > \rts$ 
			  &\cellcolor[rgb]{.95,.9,.9}send SH\_REQ to TM\newline
				\tsdesc{\msg.wts$\Leftarrow$\data.wts, \msg.pts$\Leftarrow$pts}
			  & & \cellcolor[rgb]{.8,.9,.8} 
			  & & & \cellcolor[rgb]{.7,.7,.7}
			\\\hline			
			Exclusive
			  & Hit\newline \tsdesc{pts$\Leftarrow$Max(pts, \data.wts) 
			  \newline
			  \data.rts$\Leftarrow$Max(pts, \data.rts)} & Hit \newline 
			  \tsdesc{pts$\Leftarrow$Max(pts, \data.rts+1) \newline
				\data.wts$\Leftarrow$pts \newline
				\data.rts$\Leftarrow$pts} 
			  & state$\Leftarrow$Invalid \newline
				send FLUSH\_REP to TM \newline 
				\tsdesc{\msg.wts$\Leftarrow$\data.wts, \newline
				\msg.rts$\Leftarrow$\data.rts}
			  & \cellcolor[rgb]{.7,.7,.7}
			  & \cellcolor[rgb]{.7,.7,.7}
			  & \textbf{\underline{FLUSH\_REQ}} \newline
				\tsdesc{\msg.wts$\Leftarrow$\data.wts \newline
				\msg.rts$\Leftarrow$\data.rts} \newline
				send FLUSH\_REP to TM \newline
				state$\Leftarrow$Invalid \newline
				\textbf{\underline{WB\_REQ}} \newline
				\tsdesc{\data.rts$\Leftarrow$Max(\data.rts, \data.wts+lease, req\msg.rts) \newline
				\msg.wts$\Leftarrow$\data.wts \newline
				\msg.rts$\Leftarrow$\data.rts} \newline
				send WB\_REP to TM \newline
				state$\Leftarrow$Shared
			\\\hline
		\end{tabular}
	} \\
	\end{center}
	\label{tab:prmem}
\end{table*}

\begin{table*}[ht!]
	\vspace{-.2in}
	\caption{ State Transition in Timestamp Manager.}
	\begin{center}
	{ \scriptsize
		\begin{tabular}{|p{0.5in}||p{1.3in}|p{1.1in}|p{.8in}|p{.8in}|p{1.1in}|}
			\hline
			States & SH\_REQ & EX\_REQ & Eviction & DRAM\_REP & FLUSH\_REP or \newline WB\_REP \\ \hline\hline

			Invalid 
			  & \multicolumn{2}{p{2.2in}|}{Load from DRAM}  
			  & \cellcolor[rgb]{.7,.7,.7}
			  & Fill in data \newline 
				\tsdesc{\data.wts$\Leftarrow$mts \newline
				\data.rts$\Leftarrow$mts} \newline
				state$\Leftarrow$Shared
			  & \cellcolor[rgb]{.7,.7,.7}
			\\\hline

			Shared 
			  & \tsdesc{\data.rts$\Leftarrow$Max(\data.rts, \data.wts+lease, req\msg.pts+lease)} \newline
				\textbf{\underline{if req\msg.wts=\data.wts}} \newline
				send RENEW\_REP to requester\newline
				\tsdesc{\msg.rts$\Leftarrow$\data.rts} \newline 
			   	\textbf{\underline{else}} \newline
				send SH\_REP to requester\newline
				\tsdesc{\msg.wts$\Leftarrow$\data.wts \newline
				\msg.rts$\Leftarrow$\data.rts}
  			  & \cellcolor[rgb]{.8,.9,.8} \tsdesc{\msg.rts$\Leftarrow$\data.rts} \newline
				state$\Leftarrow$Excl. \newline
				\textbf{\underline{if reqM.wts=D.wts}} \newline
				send UPGRADE\_REP to requester \newline
				\textbf{\underline{else}} \newline
				\tsdesc{\msg.wts$\Leftarrow$\data.wts} \newline
				send EX\_REP to requester
			  & \cellcolor[rgb]{.8,.9,.8}\tsdesc{mts$\Leftarrow$Max(mts, \data.rts)} 					\newline Store data to DRAM if dirty \newline
				state$\Leftarrow$Invalid
			  & \cellcolor[rgb]{.7,.7,.7}
			  & \cellcolor[rgb]{.7,.7,.7}
			\\\hline
			
			Exclusive 
			  & send WB\_REQ to the owner \newline 
				\tsdesc{\msg.rts$\Leftarrow$req\msg.pts+lease}
			  & \multicolumn{2}{c|}{send FLUSH\_REQ to the owner} 
			  & \cellcolor[rgb]{.7,.7,.7}
			  & Fill in data \newline 
				\tsdesc{\data.wts$\Leftarrow$\msg.wts, \newline
				\data.rts$\Leftarrow$\msg.rts} \newline
				state$\Leftarrow$Shared
			\\\hline
		\end{tabular}
    	}
    	\end{center}
	\label{tab:tm}
	\vspace{-.2in}
\end{table*}

In the protocol, each cacheline (denoted as $D$) has a write timestamp 
(\data.wts) and a read timestamp (\data.rts). Initially, all \pts's 
and \mts's are $1$ and all caches are empty. Some network messages 
(denoted as $M$ or $reqM$) also have timestamps associated with them.  
Each message requires at most two timestamps. 

We now discuss different cases of the Tardis protocol shown in both 
tables. 

\subsubsection{State Transition in Private Cache (\cref{tab:prmem})}

~

{\bf Load to Private Cache (column 1, 4, 5): } A load to the private 
cache is considered as a hit if the cacheline is in exclusive state or 
is in shared state and has not expired (\pts $\leq$ \rts).  Otherwise, 
a SH\_REQ is sent to the \tsman to load the data or to extend the 
existing lease. The request message has the current \wts of the 
cacheline indicating the version of the cached data.

{\bf Store to Private Cache (column 2, 4, 5): } A store to the private 
cache can only happen if the cacheline is exclusively owned by the 
core. Same as directory coherence, EX\_REQ is sent to the \tsman for 
exclusive ownership. The \rts and \wts of the private data are
updated to $Max(\pts, \rts + 1)$ because the old version might be 
loaded at timestamp \rts by another core.

{\bf Eviction (column 3): }  Evicting shared cachelines does not 
require sending any network message. The cacheline can simply be
invalidated. Evicting exclusive cachelines is the same as in directory 
coherence; the data is returned to the \tsman (through a FLUSH\_REP 
message) and the cacheline is invalidated.

{\bf Flush or Write Back (column 6): } Exclusive cachelines in the
private cache may receive flush or write back requests from the \tsman
if the cacheline is evicted from the LLC or accessed by other cores.  
A flush is handled similarly to an eviction where the data is returned 
and the line invalidated. For a write back request, the data is 
returned but the line becomes shared.  

\begin{figure*}[ht!]
	\centering
	\includegraphics[width=.95\textwidth]{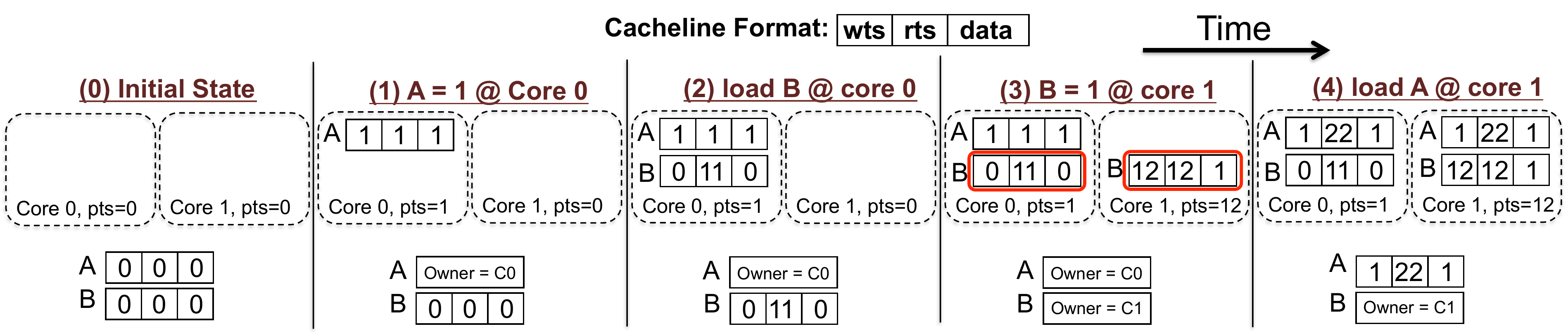}
	\vspace{-.12in}
	\caption{ An example program running with Tardis ({\it lease}$ = 
10$). Cachelines in private caches and LLC are shown. The cacheline 
format is at the top of the figure.}
	\vspace{-.2in}
\label{fig:example}
\end{figure*}

\subsubsection{State Transition in Timestamp Manager (\cref{tab:tm})}
\label{sec:state-tm}

~

{\bf Shared Request to Timestamp Manager (column 1): } If the 
cacheline is invalid in LLC, it must be loaded from DRAM. If it is 
exclusively owned by another core, then a write back request is sent 
to the owner.  When the cacheline is in the {\it Shared} state, it is 
reserved for a period of logical time by setting the \rts to be the 
end timestamp of the reservation, and the line can only be read from \wts 
to \rts in the private cache.

If the \wts of the request equals the \wts of the cacheline in the \tsman, 
the data in the private cache must be the same as the data in the LLC.  
So a RENEW\_REP is sent back to the requester without the data 
payload.  Otherwise SH\_REP is sent back with the data. 

{\bf Exclusive Request to Timestamp Manager (column 2): } An exclusive 
request can be either an exclusive load or exclusive store.  Similar 
to a directory protocol, if the cacheline is invalid, it should be 
loaded from DRAM; if the line is exclusively owned by another core, a 
flush request should be sent to the owner.  

If the requested cacheline is in shared state, however, {\em no 
invalidation messages need to be sent}. The timestamp manager can 
immediately give exclusive ownership to the requesting core 
which bumps up its local \pts to be the current \rts + 1 when it 
writes to the cacheline, i.e., jumps ahead in time.
Other cores can still read 
their local copies of the cacheline if they have not expired. This 
does not violate sequential consistency since the read operations in 
the sharing cores are ordered before the write operation in physiological time
though not necessarily in physical time. If the cacheline expires in the sharing cores, they 
will send requests to renew the line at which point they get the 
latest version of the data. 

If the \wts of the request equals the \wts of the cacheline in the 
\tsman, the data is not returned and an UPGRADE\_REP is sent to the 
requester. 

{\bf Evictions (column 3): } Evicting a cacheline in exclusive state 
is the same as in directory protocols, \ie, a flush request is sent to 
the owner before the line is invalidated. For shared cachelines, 
however, no invalidation messages are sent.  Sharing cores can still 
read their local copies until they expire -- this does not violate timestamp ordering.

{\bf DRAM (column 3, 4): } Tardis only stores timestamps on chip but 
not in DRAM. The {\it memory timestamp} (\mts) is used to maintain 
coherence for DRAM data. \mts is stored per \tsman. It indicates the 
maximal read timestamp of all the cachelines mapped to this \tsman but 
evicted to DRAM. For each cacheline evicted from the LLC, \mts is 
updated to be $Max(\rts, \mts)$. When a cacheline is loaded from DRAM, 
both its \wts and \rts are assigned to be \mts. This guarantees that 
accesses to previously cached data with the same address are ordered 
before the accesses to the cacheline just loaded from DRAM. This takes 
care of the case when a cacheline is evicted from the LLC but is still 
cached in some core's private cache. Note that multiple \mts's can be 
stored per \tsman for different ranges of cacheline addresses. In this 
paper, we only consider a single \mts per \tsman for simplicity.

\begin{table}
	\caption{ Network Messages. The check marks indicate what components the message contains. }
     	\begin{center}
	{ \scriptsize
		\begin{tabular}{|c||c|c|c|c|}
			\hline
			Message Type & \pts & \rts & \wts & data \\ \hline\hline
			
			SH\_REQ & $\surd$ & & $\surd$ & \\\hline
			EX\_REQ & & & $\surd$ & \\\hline
			FLUSH\_REQ & & & & \\\hline
			WB\_REQ & & $\surd$ & & \\\hline
			
			SH\_REP & & $\surd$ & $\surd$ & $\surd$ \\\hline
			EX\_REP & & $\surd$ & $\surd$ & $\surd$ \\\hline
			UPGRADE\_REP & & $\surd$ & & \\\hline
			RENEW\_REP & & $\surd$ & & \\\hline
			FLUSH\_REP & & $\surd$ & $\surd$ & $\surd$ \\\hline
			WB\_REP & & $\surd$ & $\surd$ & $\surd$ \\\hline

			DRAM\_ST\_REQ & & & & $\surd$ \\\hline
			DRAM\_LD\_REQ & & & & \\\hline
			DRAM\_LD\_REP & & & & $\surd$ \\\hline
		\end{tabular}
    	}
    	\end{center}
	\label{tab:msg}
	\vspace{-.3in}
\end{table}

{\bf Flush or write back response (column 5): } Finally, the flush 
response and write back response are handled in the same way as in 
directory protocols. Note that when a cacheline is exclusively owned 
by a core, only the owner has the latest \rts and \wts; the \rts and 
\wts in the \tsman are invalid and the bits can be reused to store the 
ID of the owner core.

\subsubsection{An Example Program} \label{sec:example}

We use an example to show how Tardis works with a parallel program.   
\cref{fig:example} shows how the simple program in 
Listing~\ref{lst:example} runs with the Tardis protocol. In the 
example, we assume a lease of 10 and that the instructions from Core 0 
are executed before the instructions in Core 1. 

\vspace{-.1in}
\begin{lstlisting}[language=C,label={lst:example},caption={Example 
Program}]
		   initially A = B = 0
		[Core 0]	[Core 1]
		 A = 1		 B = 1
		 print B	 print A
\end{lstlisting}

{\bf Step 1} : The store to A misses in Core 0's private cache and an 
EX\_REQ is sent to the \tsman. The store operation should happen at 
$pts=Max(pts, A.rts+1)=1$ and the A.\rts and A.\wts in the private 
cache should also be 1. The \tsman marks A as exclusively owned by 
Core 0. 

{\bf Step 2} : The load of B misses in Core 0's private cache. After 
Step 1, Core 0's \pts becomes 1. So the reservation end timestamp 
should be $Max(rts, wts+lease, pts+lease) = 11$.

{\bf Step 3} : The store to B misses in Core 1's private cache. At the 
\tsman, the exclusive ownership of B is immediately given to Core 1 at 
\pts = \rts + 1 = 12.  Note that two different versions of B exist in 
the private caches of core 0 and core 1 (marked in red circles). In 
core 0, $B = 0$ but is valid when $0 \leq$ {\it timestamp} $\leq 11$; 
in Core 1, B = 1 and is only valid when {\it timestamp} $\geq 12$.  
This does not violate sequential consistency since the loads of B at 
core 0 will be logically ordered before the loads of B at core 1, even 
if they may happen the other way around with respect to the physical 
	time.  

{\bf Step 4} : Finally the load of A misses in Core 1's private cache.  
The \tsman sends a WB\_REQ to the owner (Core 0) which updates its own 
timestamps and writes back the data. Both cores will have the same data 
with the same range of valid timestamps. 

With Tardis on sequential consistency, it is impossible for the 
example program above to output 0 for both A and B, even for 
out-of-order execution.
The reason will be discussed in \cref{sec:ooo}.

\subsection{Out-of-Order Execution} \label{sec:ooo}

So far we have assumed in-order cores, \ie, a second instruction is 
executed only after the first instruction commits and updates the 
\pts. For out-of-order cores, a memory instruction can be executed 
before previous instructions finish and thus the current \pts is not 
known. However, with sequential consistency, all instructions must 
commit in the program order. Tardis therefore enforces timestamp order 
at the commit time. 

\subsubsection{Timestamp Checking} \label{sec:ts-checking}


In the re-order buffer (ROB) of an out-of-order core, instructions 
commit in order. We slightly change the meaning of \pts to mean the 
timestamp of the last {\it committed} instruction. For sequential 
consistency, \pts still increases monotonically. Before committing an 
instruction, the timestamps are checked. 
Specifically, the following cases may happen for shared and exclusive 
cachelines, respectively. 

A shared cacheline can be accessed by load requests. And there are two 
possible cases.  

\begin{enumerate}
\item $\pts \leq \rts$. The instruction commits. $\pts \Leftarrow 
Max(\wts, \pts)$.
\item $\pts > \rts$. The instruction aborts and is restarted
with the latest \pts. Re-execution will trigger a renew request.

\end{enumerate}

An exclusive cacheline can be accessed by both load and store 
requests. And the accessing instruction can always commit with $\pts 
\Leftarrow Max(\pts, \wts)$ for a load operation and $\pts \Leftarrow 
Max(\pts, \rts+1)$ for a store operation.


There are two possible outcomes of a restarted load. If the cacheline 
is successfully renewed, the contents of the cacheline do not change.  
Otherwise, the load returns a different version of data and all the 
depending instructions in the ROB need to abort and be restarted. This 
hurts performance and wastes energy.  However, the same flushing 
operation is also required for an OoO core on a baseline directory 
protocol under the same scenario~\cite{gharachorloo1991}.  If an 
invalidation happens to a cacheline after it is executed but before it 
commits, the load is also restarted and the ROB flushed. In this case, 
the renewal failure in Tardis serves as similar functionality to an 
invalidation in directory protocols. 


\subsubsection{Out-of-Order Example}

If the example program in \cref{sec:example} runs on an out-of-order 
core, both loads may be scheduled before the corresponding stores 
making the program print $A=B=0$. In this section, we show how this 
scenario can be detected by the timestamp checking at commit time and 
thus never happens.  For the program to output $A=B=0$ in Tardis, both 
loads are executed before the corresponding stores in the timestamp 
order.
\vspace{-.05in}
\[L(A) <_{ts} S(A), \ \ \ \ \ L(B) <_{ts} S(B)\]
\vspace{-.2in}

For the instruction sequence to pass the timestamp checking, we have 
\vspace{-.05in}
\[S(A) \leq_{ts} L(B), \ \ \ \ \ S(B) \leq_{ts} L(A)\]
\vspace{-.2in}

Putting them together leads to the following contradiction.
\vspace{-.05in}
\[L(A) <_{ts} S(A) \leq_{ts} L(B) <_{ts} S(B) \leq_{ts} L(A)\]
\vspace{-.2in}

This means that at least at one core, the timestamp checking will 
fail. The load at that core is restarted with the updated \pts. The 
restarted load will not renew the lease but return the latest value 
(\ie, $1$). So at least at one core, the output value is $1$ and 
$A=B=0$ can never happen. 

\subsection{Avoiding Livelock} \label{sec:tardis-livelock}

Although Tardis strictly follows sequential consistency, it may 
generate livelock due to deferred update propagation. In 
directory coherence, a write is quickly observed by all the cores 
through the invalidation mechanism. 
In Tardis, however, a core may still read the old cached data even if 
another core has updated it, as long as the cacheline has not expired.  
In other words, the update to the locally cached data is deferred. In 
the worst case when deferment becomes indefinite, livelock occurs. For 
example, if a core spins on a variable which is set by another core, 
the \pts of the spinning core does not increase and thus the old data 
never expires. As a result, the core may spin forever without 
observing the updated data. 


We propose a very simple solution to handle this livelock problem. 
To guarantee forward progress, we only need to make sure that an update 
is {\it eventually} observed by following loads, that is, the update 
becomes globally visible within some finite physical time. This is
achieved by occasionally incrementing the \pts in each core so that 
the old data in the private cache eventually expires and the latest 
update becomes visible. 
The self increment can be periodic or based on more intelligent heuristics.
We restrict ourselves to periodic increments in this paper.


\subsection{Tardis vs. Directory Coherence} 
\label{sec:tardis-vs-dircc}

In this section, we compare Tardis to the directory coherence 
protocol. 


\subsubsection{Protocol Messages}

In \cref{tab:prmem} and \cref{tab:tm}, the advantages and 
disadvantages of Tardis compared to directory protocols are shaded in 
light green and light red, respectively. Both schemes have similar 
behavior and performance in the other state transitions (the white 
cells). 

{\bf Invalidation: } In a directory protocol, when the directory 
receives an exclusive request to a {\it Shared} cacheline, the 
directory sends invalidations to all the cores sharing the cacheline 
and waits for acknowledgements. This usually incurs significant 
latency which may hurt performance. In Tardis, however, no 
invalidation happens (\cf \cref{sec:tardis-withprmem}) and the 
exclusive ownership can be immediately returned without waiting. The 
timestamps guarantee that sequential consistency is maintained.
The elimination of invalidations makes Tardis much simpler to 
implement and reason about.

{\bf Eviction: } In a directory protocol, when a shared cacheline is 
evicted from the private cache, a message is sent to the directory 
where the sharer information is stored.  Similarly, when a shared 
cacheline is evicted from the LLC, all the copies in the private 
caches should be invalidated.  
In Tardis,  correctness does not require maintaining sharer 
information and thus no such invalidations are required. 
When a cacheline is evicted from the LLC, the copies in the private 
caches can still exist and be accessed. 

{\bf Data Renewal: } In directory coherence, a load hit only requires
the cacheline to exist in the private cache. In Tardis, however, a 
cacheline in the private cache may have expired and cannot be 
accessed. In this case, a renew request is sent to the \tsman which 
incurs extra latency and network traffic. 
In \cref{sec:opt-spec}, we present techniques to reduce the overhead 
of data renewal.

\subsubsection{Scalability}

A key advantage of Tardis over directory coherence protocols is 
scalability.  Tardis only requires the storage of timestamps for each 
cacheline and the owner ID for each LLC cacheline ($O(\log{N})$, where 
$N$ is the number of cores). In practice, the same hardware bits can 
be used for both timestamps and owner ID in the LLC; because when the 
owner ID needs to be stored, the cacheline is exclusively owned and 
the \tsman does not maintain the timestamps. 

On the contrary, a canonical directory coherence protocol maintains 
the list of cores sharing a cacheline which requires $O(N)$ storage 
overhead. Previous works proposed techniques to improve the 
scalability of directory protocols by introducing broadcast or other 
complexity. 
They are discussed in \cref{sec:related-scale-dircc}.


\subsubsection{Simplicity}

Another advantage of Tardis is its conceptual simplicity and elegance.  
Tardis is directly derived from the definition of sequential 
consistency and the timestamps explicitly express the global memory 
order. This makes it easier to argue the correctness of the protocol.  
Concretely, given that Tardis does not have to multicast/broadcast 
invalidations and collect
acknowledgements, the number of transient states in Tardis is smaller
than that of a directory protocol.

\section{Optimizations and Extensions} \label{sec:opt}

We introduce several optimizations in the Tardis protocol in this 
section, which were enabled during our evaluation. The evaluation
of the extensions described here is deferred to future work.

\subsection{Speculative Execution} \label{sec:opt-spec}

As discussed in \cref{sec:tardis-vs-dircc}, the main disadvantage of 
Tardis compared to directory coherence protocol is the renew request.  
In a pathological case, the \pts of a core may rapidly increase since 
some cachelines are frequently read-write shared by different cores.  
Meanwhile, the read-only cachelines will frequently expire and a large 
number of renew requests are generated incurring both latency and 
network traffic. Observe, however, that most renew requests will 
successfully extend the lease and the renew response does not transfer 
the data.  This significantly reduces the network traffic of renewals.  
More important, this means that the data in the expired cacheline is 
actually valid and we could have used the value without stalling the 
pipeline of the core. Based on this observation, we propose the use of 
speculation to hide renew latency. When a core reads a cacheline which 
has expired in the private cache, instead of stalling and waiting for 
the renew response, the core reads the current value and continues 
executing speculatively. If the renewal fails and the latest cacheline 
is returned, the core rolls back by discarding the speculative 
computation that depends on the load. The rollback process is almost 
the same as a branch misprediction which has already been implemented 
in most processors.


For processors that can buffer multiple uncommitted instructions,
successful renewals (which is the common case) do not hurt 
performance. Speculation failure does incur performance overhead since 
we have to rollback and rerun the instructions speculatively executed.  
However, if the same instruction sequence is executed in a directory 
protocol, the expired cacheline should not be in the private cache in 
the first place; the update from another core should have already 
invalidated this cacheline and a cache miss should happen. As a 
result, in both Tardis and directory coherence, the value of the load 
should be returned at the same time incurring the same latency and 
network traffic. Tardis still has some extra overhead as it needs to 
discard the speculated computation, but this overhead is relatively 
small. 

Speculation successfully hides renew latency in most cases. The renew 
messages, however, may still increase the on-chip network traffic.  
This is especially problematic if the private caches have a large number
of {\em shared} cachelines that all expire when the \pts jumps ahead due to a 
write or self increment. This is a fundamental disadvantage of Tardis 
compared to directory coherence protocols. According to our 
evaluations in \cref{sec:eval}, however, Tardis has good performance
and acceptable network overhead on real benchmarks even with this 
disadvantage. We leave solutions to pathologically bad cases to future work.  



\subsection{Timestamp Compression} \label{sec:opt-compress}

In Tardis, all the timestamps increase monotonically and may roll 
over. One simple solution is to use 64-bit timestamps which never roll 
over in practice. This requires 128 extra bits to be stored per 
cacheline, which is a significant overhead.  Observe, however, that 
the higher bits in a timestamp change infrequently and are usually the 
same across most of the timestamps. We exploit this observation and 
propose to compress this redundant information using a {\it 
base-delta} compression scheme.  

In each cache, a {\it base timestamp} (\bts) stores the common high 
bits of \wts and \rts. In each cacheline, only the {\it delta 
timestamps} (\dts) are stored (\dwts = \wts$-$ \bts and \drts = 
\rts$-$ \bts).  The actual timestamp is the sum of the \bts and the 
corresponding \dts. The \bts is 64 bits to prevent rollover; and there 
is only one \bts per cache. The per cacheline \dts is much shorter to 
reduce the storage overhead.  

When any \dts in the cache rolls over, we will rebase where the local 
\bts is increased and all the \dts in the cache are decreased by the 
same amount, \ie, half of the maximum \dts. For simplicity, we assume 
that the cache does not serve any request during the rebase operation. 

Note that increasing the \bts in a cache may end up with some \dts 
being negative. In this case, we just set the \dts to $0$. This 
effectively increases the \wts and \rts in the cacheline but it does 
not violate the consistency model. Consider a shared LLC cacheline or 
an exclusive private cacheline -- the \wts and \rts can be safely 
increased. Increasing the \wts corresponds to writing the same data to 
the line at a later logical time, and increasing the \rts corresponds 
to a hypothetical read at a later logical time. Neither operation 
violates the rules of sequential consistency. Similarly, for a shared 
cacheline in the private cache, \wts can be safely increased as long 
as it is smaller than \rts. However, \rts can not be increased without 
coordinating with the \tsman. So if \drts goes negative in a shared 
line in a private cache, we simply invalidate the line from the cache.  
The last possible case is an exclusive cacheline in the LLC.  No 
special operation is required since the \tsman neither has the 
timestamps nor the data in this case.

The key advantage of this base-delta compression scheme is that all 
computation is local to each cache without coordination between 
different caches.  This makes the scheme very scalable.

It is possible to extend the base-delta scheme to \wts and \rts to 
further compress the timestamp storage. Specifically, \wts can be 
treated as the \bts and we only need to store the $\drts = \rts - 
\wts$ which can be even shorter than $\rts - \bts$. We defer an
evaluation of this extension to future work.

The scheme discussed here does not compress the timestamps over the 
network and we assume that the network messages still use 64-bit 
timestamps. It is possible to reduce this overhead by extending the 
base-delta scheme over the network but this requires coordination 
amongst multiple caches. We did not implement this extension in order 
to keep the basic protocol simple and straightforward.


\subsection{Private Write}
\label{sec:private-write}

According to \cref{tab:prmem}, writing to a cacheline in exclusive 
state updates both \wts and \pts to $Max(\pts, \rts + 1)$. If the core 
keeps writing to the same address, the \pts will keep increasing 
causing other cachelines to expire. If the updated cacheline is 
completely private to the updating thread, however, there is actually 
no need to increase the timestamps in order to achieve sequential 
consistency.  According to our definition of global memory order 
(Definition~\ref{def:mem-order}), we can use physical time to order 
these operations implicitly without increasing the \pts. 

Specifically, when a core writes to a cacheline, the \textit{modified 
bit} will be set.  For future writes to the same cacheline, if the bit 
is set, then the \pts, \wts and \rts are just set to $Max(\pts, 
\rts)$. This means that \pts will not increase if the line is 
repeatedly written to. The optimization will significantly reduce the 
rate at which timestamps increase if most of the accesses from a core 
are to thread private data.

This optimization does not violate sequential consistency because 
these writes with the same timestamp are ordered correctly in the 
physical time order and thus they are properly ordered in the global 
memory order.  

\subsection{Extension: Exclusive and Owned States}
\label{sec:estate}

In this paper, {\it MSI} has been used as the baseline directory 
coherence protocol.  {\it MSI} is the simplest protocol and optimized 
ones require {\bf E} (Exclusive) and/or {\bf O} (Owned) states. The 
resulting protocols are {\it MESI}, {\it MOSI} and {\it MOESI}. In 
this section, we show that Tardis is compatible with both {\it E} and 
{\it O} states.

Similar to the {\it M} state, the {\it E} state allows the cacheline
to be exclusively cached upon a SH\_REQ if no other sharers exist.
The core having the cacheline can update the data silently without
notifying the directory. In the directory, {\it M} or {\it E}
cachelines are handled in the same way; an invalidation is sent to the
core exclusively caching the line if another core requests it. In 
Tardis, we can support the {\it E}
state by always returning a cacheline in exclusive state if no other
cores seem to be caching it.  Note that even if other cores are
sharing the line, it can still be returned to the requester in
exclusive state. The argument for this is similar to the write after
shared argument in \cref{sec:state-tm}; \ie, the lines shared and the
line exclusively cached have different ranges of valid timestamps.
However, this may not be best for performance. Therefore, we would 
like to return a cacheline in exclusive state only if the cacheline 
{\em seems} to be private. We can add an extra bit for
each cacheline indicating whether any core has accessed it since it
was put into the LLC. And if the bit is unset, the requesting core
gets the line in exclusive state, else in shared state.
{\it E} states will reduce the number of renewals required since 
cachelines in {\it E} state will not expire.

The {\it O} state allows a cacheline to be dirty but shared in the 
private caches. Upon receiving the WB\_REQ request, instead of writing 
the data back to the LLC or DRAM, the core can change the cacheline to 
{\it O} state and directly forward the data to the requesting core. In 
Tardis, the {\it O} state can also be supported by keeping track of 
the owner at the \tsman. SH\_REQs to the \tsman are forwarded to the 
owner which does cache-to-cache data transfers.  Similar to the basic 
Tardis protocol, when the owner is evicted from the private cache, the 
cacheline is written back to the LLC or the DRAM and its state in the 
\tsman is changed to Shared or Invalid.  

\begin{figure*}[ht!]
	\centering
	\includegraphics[width=.95\textwidth]{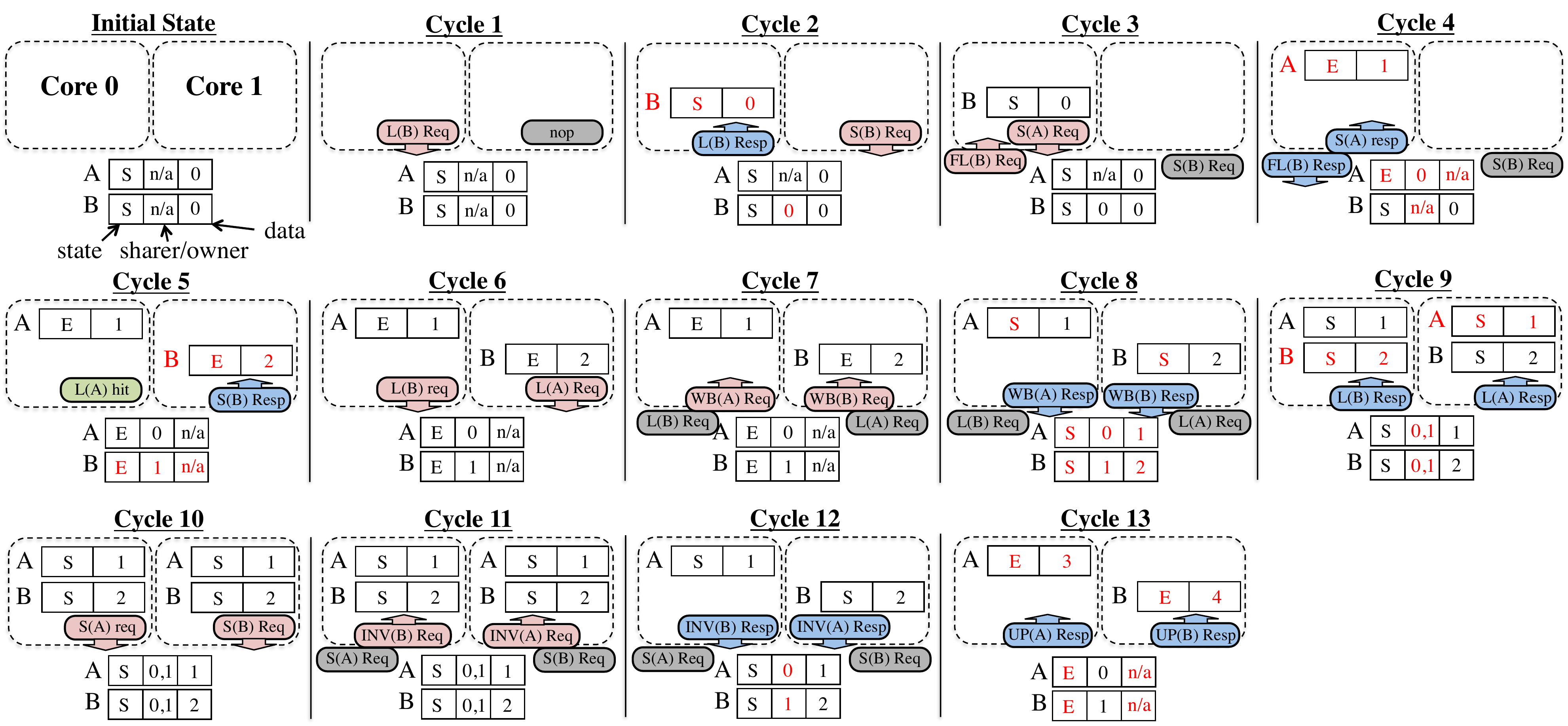}
	\caption{ Execution of the case study program with a directory 
	coherence protocol.  }
	\label{fig:case-directory}
\end{figure*}

\begin{figure*}[ht!]
	\centering
	\includegraphics[width=.95\textwidth]{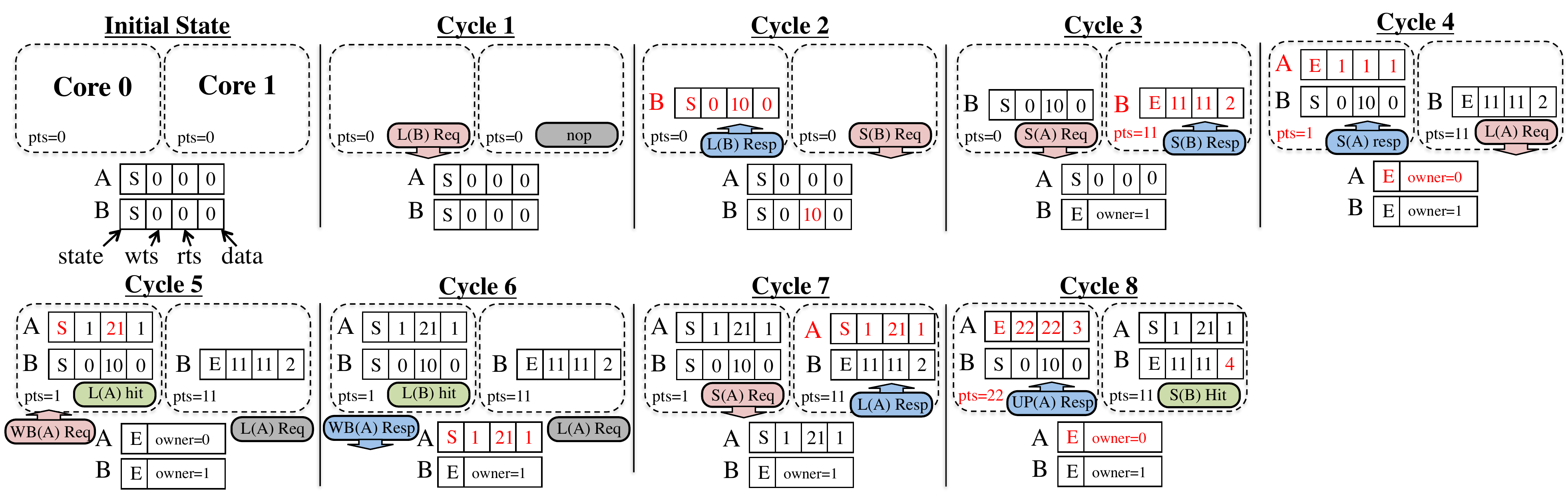}
	\caption{ Execution of the case study program with Tardis 
	protocol.  }
	\label{fig:case-tardis}
\end{figure*}

\subsection{Extension: Remote Word Access}

Traditionally, a core loads a cacheline into the private cache before 
accessing the data. But it is also possible to access the data 
remotely without caching it. Remote word access has been studied in 
the context of locality-aware directory coherence~\cite{kurian2013}.  
Remote atomic operation has been implemented on Tilera 
processors~\cite{david2013, hoffmann2010}.  Allowing data accesses or 
computations to happen remotely can reduce the coherence messages and 
thus improve performance~\cite{yu2014}.

However, it is not easy to maintain the performance gain of these 
remote operations with directory coherence under TSO or sequential 
consistency. For a remote load operation (which might be part of a 
remote atomic operation), it is not very easy to determine its global 
memory order since it is hard to know the physical time when the load 
operation actually happens. As a result, integration of remote access 
with directory coherence is possible but fairly involved 
\cite{kurian2014}.

Consider the example program in Listing~\ref{lst:example} where all 
memory requests are remote accesses. If all requests are issued 
simultaneously, then both loads may be executed before both stores and 
the program outputs $A = B = 0$. It is not easy to detect this 
violation in a directory protocol since we do not know when each 
memory operation happens. As a result, either the remote accesses are 
sequentially issued or additional mechanisms need to be 
added~\cite{kurian2014}. 

In Tardis, however, memory operations are ordered through timestamps.  
It is very easy to determine the memory order for a remote access 
since it is simply the timestamp of the operation. In Tardis, multiple 
remote accesses can be issued in parallel and the order can be checked 
after they return. If any load violates the memory order, it can be 
reissued with the updated timestamp information (similar to timestamp 
checking in an out-of-order core).

\subsection{Other Extensions}

Atomic operations in Tardis can be implemented the same way as in 
directory protocols. Tardis can be extended to relaxed consistency 
models such as Total Store Order (TSO)
implemented in Intel x86 processors~\cite{sewell2010}.
Tardis can work with both private Last Level Cache (LLC) or shared LLC.

\section{Case Study} \label{sec:case}

In this section, we use an example parallel program as a case study to 
compare Tardis with an MSI directory protocol. 

\subsection{Example}

Listing~\ref{lst:case} shows the parallel program we use for the case 
study. In this program, the two cores issue loads and stores to 
addresses \texttt{A} and \texttt{B}. The \texttt{nop} in Core 1 means 
that the core spends that cycle without accessing the memory subsystem.  
The program we use here is a contrived example to highlight the 
difference between Tardis and a directory coherence protocol.

\begin{lstlisting}[language=C,label={lst:case},caption={The case study
parallel program}]
		[Core 0]	[Core 1]
		 L(B)		 nop
		 A = 1		 B = 2
		 L(A)		 L(A)
		 L(B)		 B = 4
		 A = 3
\end{lstlisting}

\cref{fig:case-directory} shows the execution of the program on a 
directory coherence protocol and \cref{fig:case-tardis} shows how it 
is executed on Tardis. A cacheline is either in shared (\textit{S}) or 
exclusive (\textit{E}) state. For Tardis, a static lease of 10 is 
used. Initially, all private caches are empty and all timestamps are 0.  
We will explain step by step how Tardis executes the program and 
highlight the differences between Tardis and the directory protocol.

\textbf{Cycle 1 and 2}: Core 0 sends a shared request to address 
\textit{B} in cycle 1, and receives the response in cycle 2. The 
cacheline is reserved till timestamp 10. Core 1 sends an exclusive 
request to address \textit{B} at cycle 2. In these two cycles, both 
the directory protocol and Tardis have the same network messages sent and 
received.

\textbf{Cycle 3}: In Tardis, the exclusive request from core 1 sees 
that address \textit{B} is shared till timestamp 10. The exclusive 
ownership is instantly returned and the store is performed at 
timestamp 11. In the directory protocol, however, an invalidation must 
be sent to Core 0 to invalidate address \textit{B}. As a result, the 
exclusive response is delayed to cycle 5. At this cycle, core 0 
sends an exclusive request to address \textit{A}.

\textbf{Cycle 4}: In both Tardis and the directory protocol, address 
\textit{A}'s exclusive ownership can be instantly returned to core 0 
since no core is sharing it. The \pts of core 0 becomes 1 after 
performing the store. Core 1 performs a shared request to address 
\textit{A} which needs to get the latest data from core 0 through 
write back. So the shared response returns in cycle 7.  The same 
\textit{L(A)} instruction in the directory protocol incurs the same latency and 
network traffic from cycle 6 to 9.

\textbf{Cycle 5 and 6}: In cycle 5, the \textit{L(A)} instruction in 
core 0 hits in the private cache and thus no request is sent. Also in 
core 0, the write back request increases address \textit{A}'s \rts to 
21 since the requester's (core 1) \pts is 11 and the lease is 10. In 
cycle 6, the \textit{L(B)} instruction in core 0 hits in the private 
cache since the \pts is 1 and the cached address \textit{B} is valid 
till timestamp 10. In the directory protocol, the same \textit{L(B)} 
instruction is also issued at cycle 6. However, it misses in the 
private cache since the cacheline was already invalidated by core 1 at 
cycle 4. So a write back request to core 1 needs to be sent and the 
shared response returns at cycle 9.

\textbf{Cycle 7 and 8}: At cycle 7, core 0 sends an exclusive request 
to address \textit{A} and core 1 gets the shared response to address 
\textit{A}. At cycle 8, the exclusive ownership of address \textit{A} 
is instantly returned to core 0 and the store happens at timestamp 22 
(because addresss \textit{A} has been reserved for reading until 
timestamp 21). In the directory protocol, the same \textit{S(A)} 
instruction happens at cycle 10 and the shared copy in core 1 must be 
invalidated before exclusive ownership is given. Therefore, the 
exclusive response is returned at cycle 13. Also in cycle 8 in Tardis, 
core 1 stores to address \textit{B}. The store hits in the private cache.  
In the directory protocol, the same store instruction happens at cycle 
10. Since core 0 has a shared copy of address \textit{B}, an 
invalidation must be sent and the exclusive response is returned at 
cycle 13. 

\subsection{Discussion}

In this case study, the cycle saving of Tardis mainly comes from the 
removal of invalidations. While a directory protocol requires that 
only one version of an address exist at any point in time across all 
caches, Tardis allows multiple versions to coexist as long as they 
are accessed at different timestamps. 

The \pts in each core shows how Tardis orders the memory operations.  
At cycle 3, core 1's \pts jumps to 11. Later at cycle 4, core 0's \pts 
jumps to 1. Although the operation from core 0 happens later than the 
operation from core 1 in physical time, it is the opposite in global 
memory and physiological time order. Later at cycle 8, core 0's \pts jumps to 
22 and becomes bigger than core 1's \pts.

In Tardis, a load may still return a old version of an address after 
it is updated by a different core, as long as sequential consistency 
is not violated. As a result, Tardis may produce a different 
instruction interleaving than a directory protocol.  
Listings~\ref{lst:inter-dir} and \ref{lst:inter-tardis} show the 
instruction interleaving of the directory protocol and Tardis,
respectively, on our example program.

\noindent
\hspace{.1in}
\begin{minipage}[t]{.45\columnwidth}
\begin{lstlisting}[language=C,label={lst:inter-dir},caption={Instruction 
interleaving in directory protocol}, escapeinside={@}{@}]

 [Core 0]     [Core 1]

  L(B)   

  A = 1        B = 2

  L(A)         L(A)
  
  L(B)         B = 4
  
  A = 3
@
\begin{tikzpicture}[overlay,remember picture]
\node (A) at (1, 2.9) {};
\node (B) at (2.5, 2.4) {};
\node [rotate=-20] (L1) at (1.8, 2.8) {WAR};
\draw[->] (A) to node {} (B);

\node (C) at (2.5, 2.3) {};
\node (D) at (1, 1.1) {};
\node [rotate=38] (L2) at (1.7, 1.9) {RAW};
\draw [->] (C) to node {} (D);

\node (E) at (1, 1) {};
\node (F) at (2.5, 1) {};
\node (L3) at (1.55, 1.15) {WAR};
\draw [->] (E) to node {} (F);

\node (G) at (2.5, 1.6) {};
\node (H) at (1.2, 0.3) {};
\node [rotate=45] (L3) at (2.1, 1.4) {WAR};
\draw [->] (G) to node {} (H);
\end{tikzpicture}
@
\end{lstlisting}
\end{minipage}
\hfill
\begin{minipage}[t]{.45\columnwidth}
\begin{lstlisting}[language=C,label={lst:inter-tardis},caption={Instruction 
interleaving in Tardis}, escapeinside={@}{@}]
 
 [Core 0]     [Core 1]

  L(B)   

  A = 1        B = 2

  L(A)         L(A)
  
  L(B)         B = 4
  
  A = 3
@
\begin{tikzpicture}[overlay,remember picture]
\node (A) at (1, 2.9) {};
\node (B) at (2.5, 2.4) {};
\node [rotate=-20] (L1) at (1.8, 2.8) {WAR};
\draw[->] (A) to node {} (B);

\node (C) at (2.5, 2.3) {};
\node (D) at (1, 1.1) {};
\node [rotate=38] (L2) at (1.7, 1.9) {WAR};
\draw [->] (D) to node {} (C);

\node (G) at (2.5, 1.6) {};
\node (H) at (1.2, 0.3) {};
\node [rotate=45] (L3) at (2.1, 1.4) {WAR};
\draw [->] (G) to node {} (H);
\end{tikzpicture}
@
\end{lstlisting}
\end{minipage}

In the directory protocol, the second \textit{L(B)} instruction from 
core 0 is between the two stores to address \textit{B} from core 1 in 
the global memory order. In Tardis, however, the same \textit{L(B)} 
instruction is ordered before both stores. Such reordering is possible 
because Tardis enforces sequential consistency in physiological time 
order which can be different from physical time order.

\section{Evaluation} \label{sec:eval}

We now evaluate the performance of Tardis in the context of multi-core 
processors.

\subsection{Methodology}

We use the Graphite~\cite{graphite} multi-core simulator for our 
experiments. The default hardware parameters are listed in 
\cref{tab:system}. The simplest directory protocol {\it MSI} is used 
as the baseline in this section.\footnote{Other states, e.g., {\bf O} 
(Owner) and {\bf E} (Exclusive) can be added to an MSI protocol
to improve performance; such states can be added to Tardis as well
to improve performance as described in \cref{sec:estate}.} This baseline keeps the full sharer 
information for each cacheline and thus incurs non-scalable storage 
overhead. To model a more scalable protocol, we use the
Ackwise~\cite{ATAC} protocol which keeps a limited number of sharers 
and broadcasts invalidations to all cores when the number of sharers exceeds the limit.

 
\begin{table}
    \caption{ System Configuration. }
	\begin{center}
	{ \scriptsize
        \begin{tabular}{|l|l|}
            \hline
			\multicolumn{2}{|c|}{System Configuration} \\
            \hline
			Number of Cores				& N = 64 @ 1~GHz\\
			Core Model                  & In-order, Single-issue \\
			\hline
            \hline
			\multicolumn{2}{|c|}{Memory Subsystem} \\			
            \hline
			Cacheline Size        		& 64~bytes \\
			L1 I Cache					& 16~KB, 4-way \\
			L1 D Cache                	& 32~KB, 4-way \\
			Shared L2 Cache per Core    & 256~KB, 8-way \\
			DRAM Bandwidth				& 8 MCs, 10~GB/s per MC\\
			DRAM Latency				& 100~ns\\
			\hline
			\hline
			\multicolumn{2}{|c|}{2-D Mesh with XY Routing} \\
			\hline 
			Hop Latency					& 2 cycles (1-router,
										1-link)\\
			Flit Width					& 128 bits\\
			\hline
			\hline
			\multicolumn{2}{|c|}{Tardis Parameters} \\
			\hline
			Lease						& 10\\
			Self Increment Period		& 100 cache accesses\\
			Delta Timestamp Size		& 20~bits\\
			L1 Rebase Overhead 			& 128~ns\\
			L2 Rebase Overhead			& 1024~ns \\
			\hline
		\end{tabular}
    }
	\end{center}
    \label{tab:system}
	\vspace{-.2in}
\end{table}

In our simulation mode, Graphite includes functional correctness 
checks, where the simulation fails, e.g., if wrong values are read.  
All the benchmarks we evaluated in this section completed which 
corresponds to a level of validation of Tardis and its Graphite 
implementation. Formal verification of Tardis can be found in 
\cite{tardis-proof}.

Splash-2~\cite{splash2} benchmarks are used for performance 
evaluation. For each experiment, we report both the throughput (in 
bars) and network traffic (in red dots).

\subsubsection{Tardis Configurations}

\cref{tab:system} also shows the default Tardis configuration. For 
load requests, a static lease of 10 has been chosen. The \pts at each 
core self increments by one for every 100 cache accesses (self 
increment period).  The Base-delta compression scheme is applied with 
20-bit delta timestamp size and 64-bit base timestamp size.  When the 
timestamp rolls over, the rebase overhead is $128~ns$ in L1 and 
$1024~ns$ in an LLC slice.

Static lease and self increment period are chosen in this paper for 
simplicity -- both parameters can be dynamically changed for better 
performance based on the data access pattern. Exploration of such 
techniques is left for future work.   

\subsection{Main Results}
\label{sec:performance-study}

\begin{figure}[t]
	\centering
	\includegraphics[width=\columnwidth]{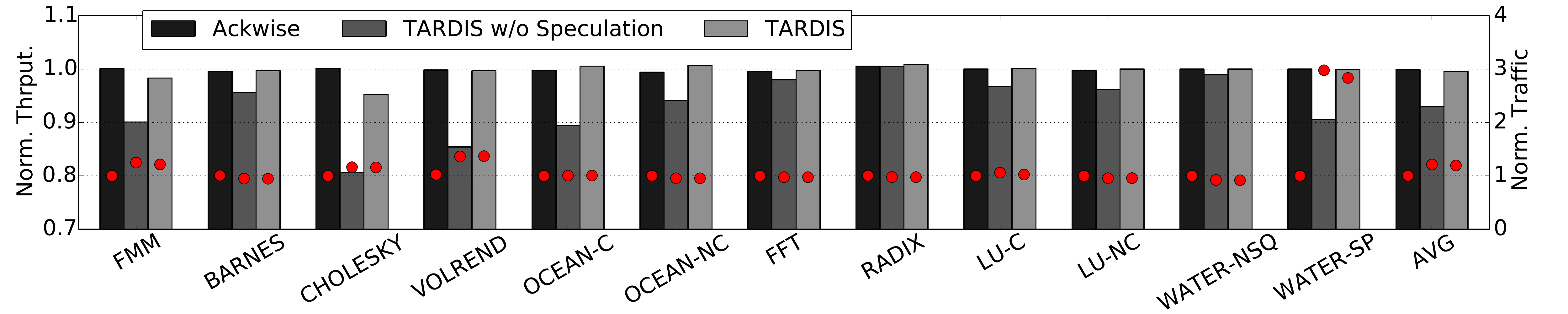}
	\vspace{-.3in}
	\caption{ Performance of Tardis at 64 cores. Both throughput
	(bars) and network traffic (dots) are normalized to baseline MSI.}
	\vspace{-.1in}
	\label{fig:main}
\end{figure}

\begin{figure}[t]
	\centering
	\includegraphics[width=\columnwidth]{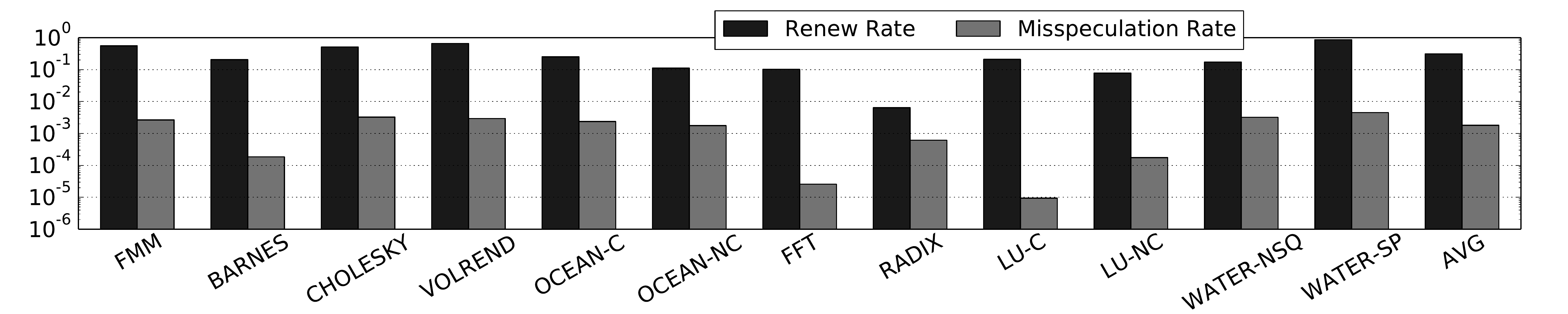}
	\vspace{-.3in}
	\caption{ Renew and misspeculation rate of Tardis at 64 cores.
	Y-axis is in log scale.}
	\vspace{-.1in}
	\label{fig:renew}
\end{figure}

\subsubsection{Throughput}

\cref{fig:main} shows the throughput of Ackwise and Tardis on 64 
in-order cores, normalized to baseline MSI. For Tardis, we also show 
the performance with speculation turned off. For most benchmarks, 
Tardis achieves similar performance compared to the directory 
baselines. On average, the performance of Tardis is within 0.5\% of 
the baseline MSI and Ackwise.   

If the speculation is turned off, Tardis's performance becomes 7\% 
worse than MSI. In this case, the core stalls while waiting
for the renewal, in contrast to the default Tardis where the core
reads the value speculatively and continues execution. Since most 
renewals are successful, speculation hides a significant amount of
latency and makes a big difference in performance.  

\subsubsection{Network Traffic}

The red dots in \cref{fig:main} show the network traffic of Ackwise 
and Tardis normalized to the baseline MSI. On average, Tardis with and 
without speculation incurs 19.4\% and 21.2\% more network traffic.  
Most of this traffic comes from renewals.
\cref{fig:renew} shows the percentage of renew requests and 
missspeculations out of all LLC accesses. Note that the y-axis is in 
log scale.

In benchmarks with lots of synchronizations (\eg, {\tt CHOLESKY},  
{\tt VOLREND}), cachelines in the private cache frequently expire 
generating a large number of renewals. In {\tt VOLREND}, for example, 
65.8\% of LLC requests are renew requests which is 2$\times$ of normal 
LLC requests. As discussed in \cref{sec:tardis-vs-dircc}, a successful 
renewal only requires a single flit message which is cheaper than a 
normal LLC access. So the relative network traffic overhead is small 
(36.9\% in {\tt VOLREND} compared to baseline MSI).

An outlier is {\tt WATER-SP}, where Tardis increases the network 
traffic by 3$\times$.  This benchmark has very low L1 miss rate and 
thus very low network utilization. Even though Tardis incurs 3$\times$ 
more traffic, the absolute amount of traffic is still very small. 

In many other benchmarks (\eg, {\tt BARNES}, {\tt WATER-NSQ}, etc.), 
Tardis has less network traffic than baseline MSI. The traffic 
reduction comes from the elimination of invalidation and cache 
eviction traffic.

From \cref{fig:renew}, we see that the misspeculation rate for Tardis 
is very low, less than 1\% renewals failed on average. A speculative 
load is considered a miss if the renew fails and a new version of data 
is returned. Having a low misspeculation rate indicates that the vast 
majority of renewals are successful.  


\subsubsection{Timestamp Discussion}

\begin{table}
	\caption{ Timestamp Statistics }
	\begin{center}
	{ \scriptsize
		\begin{tabular}{|c|c|c|}
			\hline Benchmarks & Ts. Incr. Rate & Self Incr.  Perc.  \\
			 & (cycle / timestamp) & \\
			\hline
			\hline
			FMM  &  322  &  22.5\%  \\
			\hline
			BARNES  &  155  &  33.7\%  \\
			\hline
			CHOLESKY  &  146  &  35.6\%  \\
			\hline
			VOLREND  &  121  &  23.6\%  \\
			\hline
			OCEAN-C  &  81  &  7.0\%  \\
			\hline
			OCEAN-NC  &  85  &  5.6\%  \\
			\hline
			FFT  &  699  &  88.5\%  \\
			\hline
			RADIX  &  639  &  59.3\%  \\
			\hline
			LU-C  &  422  &  1.4\%  \\
			\hline
			LU-NC  &  61  &  0.1\%  \\
			\hline
			WATER-NSQ  &  73  &  12.8\%  \\
			\hline
			WATER-SP  &  363  &  29.1\%  \\
			\hline \hline
			AVG 	  & 263   & 26.6\% \\
			\hline
			\end{tabular}
    }
	\end{center}
	\label{tab:ts-stats}
	\vspace{-.2in}
\end{table}

\cref{tab:ts-stats} shows how fast the \pts in a core increases, in 
terms of the average number of cycles to increase the \pts by 1.  
\cref{tab:ts-stats} also shows the percentage of \pts increasing 
caused by self increment (cf. \cref{sec:tardis-livelock}).

Over all the benchmarks, \pts is incremented by 1 every 263 cycles.  
For a delta timestamp size of 20 bits, it rolls over every 0.28 seconds. In 
comparison, the rebase overhead (128 ns in L1 and 1 $\mu$s in L2) becomes 
negligible.  This result also indicates that timestamps in Tardis 
increase very slowly. This is because they can only be increased from 
accessing shared read/write cachelines or self increment. 
 
On average, 26.6\% of timestamp increasing is caused by self increment 
and the percentage can be as high as 88.5\% (\texttt{FFT}). This has 
negative impact on performance and network traffic since unnecessarily 
increasing timestamps causes increased expiration and renewals. Better 
livelock avoidance algorithms can resolve this issue;
we leave this for future work.

\subsection{Sensitivity Study}

\subsubsection{In-order vs. Out-of-Order Core}

\begin{figure}[t]
	\centering
	\includegraphics[width=\columnwidth]{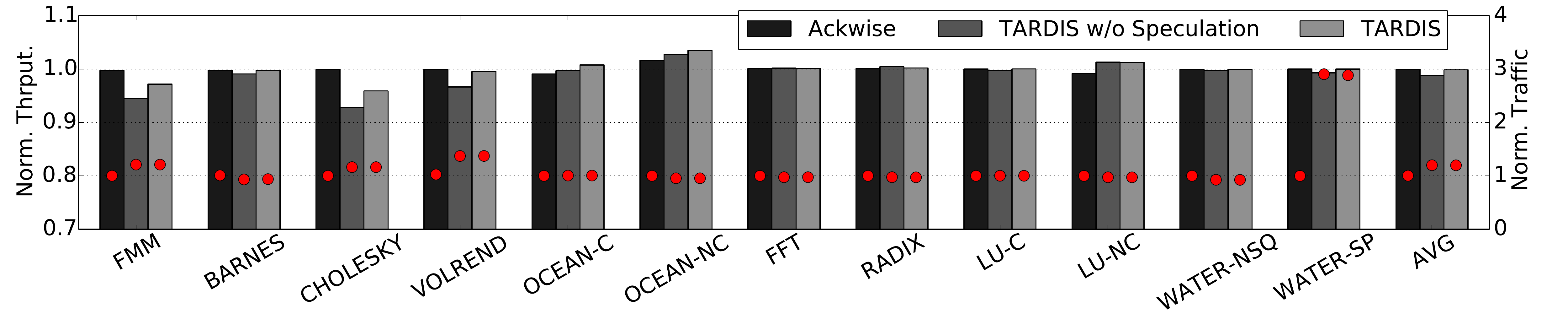}
	\vspace{-.3in}
	\caption{ Performance of Tardis on 64 out-of-order cores.}
	\vspace{-.1in}
	\label{fig:ooo}
\end{figure}

\cref{fig:ooo} shows the performance of Tardis on out-of-order cores.
Compared to in-order cores (\cref{fig:main}), the performance impact
of speculation is much smaller. When a renew request is outstanding,
an out-of-order core is able to execute independent instructions even
if it does not speculate. As a result, the renewal's latency can still 
be hidden. On average, Tardis with and without speculation is 0.2\% 
and 1.2\% within the performance of baseline MSI respectively.  

The normalized traffic of Tardis on out-of-order cores is not much 
different from in-order cores. This is because both core models follow 
sequential consistency and the timestamps assigned to the memory 
operations are virtually identical. As a result, the same amount of 
renewals is generated.  

\subsubsection{Self Increment Period} \label{sec:eval-period}

\begin{figure}[t]
	\centering
	\includegraphics[width=\columnwidth]{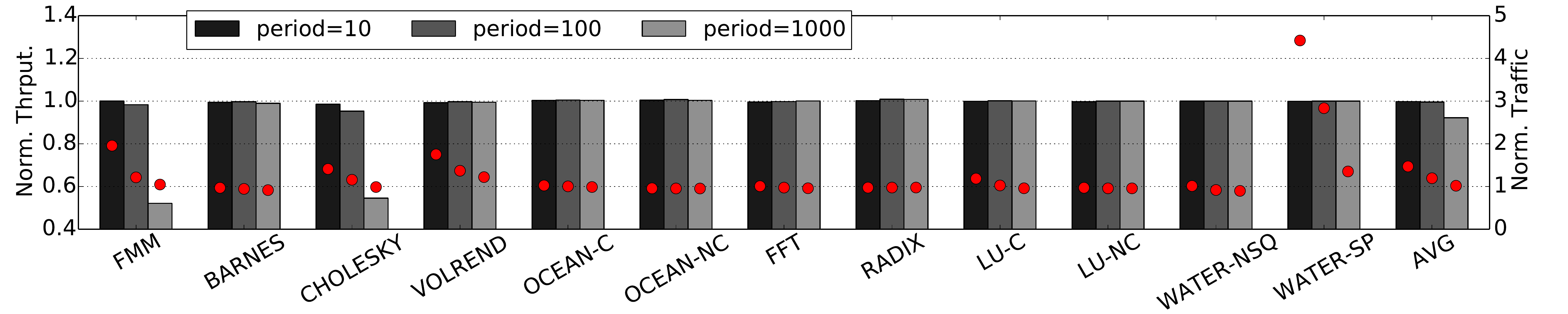}
	\vspace{-.3in}
	\caption{ Performance of Tardis with different self increment
	period.}
	\vspace{-.1in}
	\label{fig:period}
\end{figure}

As discussed in \cref{sec:tardis-livelock}, we periodically increment 
the \pts at each core to avoid livelock. The {\it self increment 
period} specifies the number of data cache accesses before self 
incrementing the \pts by one. If the period is too small, the \pts 
increases too fast causing more expirations; more renewals will be 
generated which increases network traffic and hurts performance. Fast 
growing \pts's also overflow the \wts and \rts more frequently (\cf 
\cref{sec:tssize}) which also hurts performance. If the period is too 
large, however, an update at a remote core may not be observed locally 
quickly enough, which degrades performance.  

\cref{fig:period} shows the performance of Tardis with different self 
increment period. The performance of most benchmarks is not sensitive 
to this parameter. In {\tt FMM} and {\tt CHOLESKY}, performance goes 
down when the period is 1000. This is because these two benchmarks 
heavily use spinning (busy waiting) to synchronize between threads. If 
the period is too large, the core spends a long time spinning on the 
stale value in the private cache and cannot make forward progress. 

Having a larger self increment period always reduces the total network 
traffic because of fewer renewals. Given the same performance, a 
larger period should be preferred due to network traffic reduction.  
Our default self increment period is 100 which has reasonable 
performance and network traffic.  

Ideally, the self increment period should dynamically adjust to the 
program's behavior. For example, the period can be smaller during 
spinning but larger for the rest of the program where there is no need 
to synchronize. Exploration of such schemes is deferred to future 
work.  

\subsubsection{Scalability}

\begin{figure}[t]
	\centering
	\subfloat[16 Cores]{
		\includegraphics[width=\columnwidth]{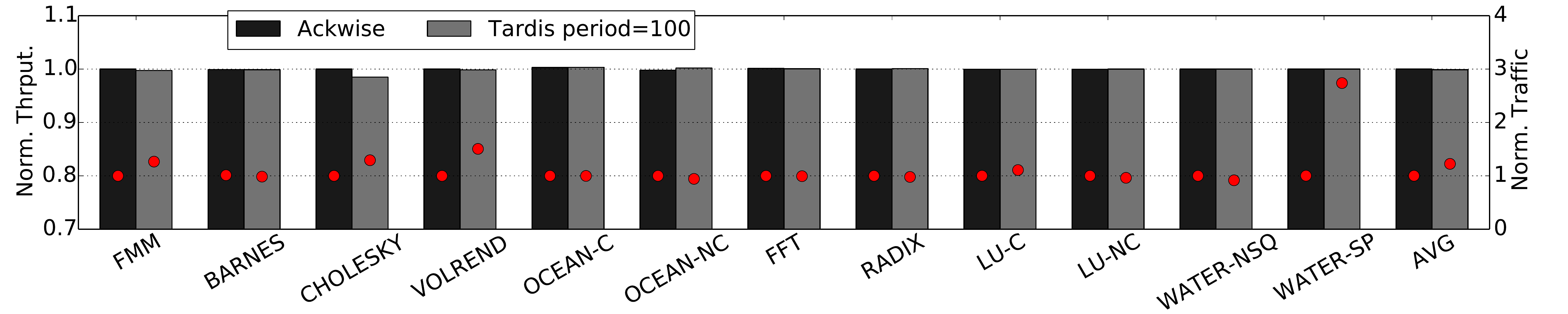}
	} \\ \vspace{-.1in}
	\subfloat[256 Cores]{
		\includegraphics[width=\columnwidth]{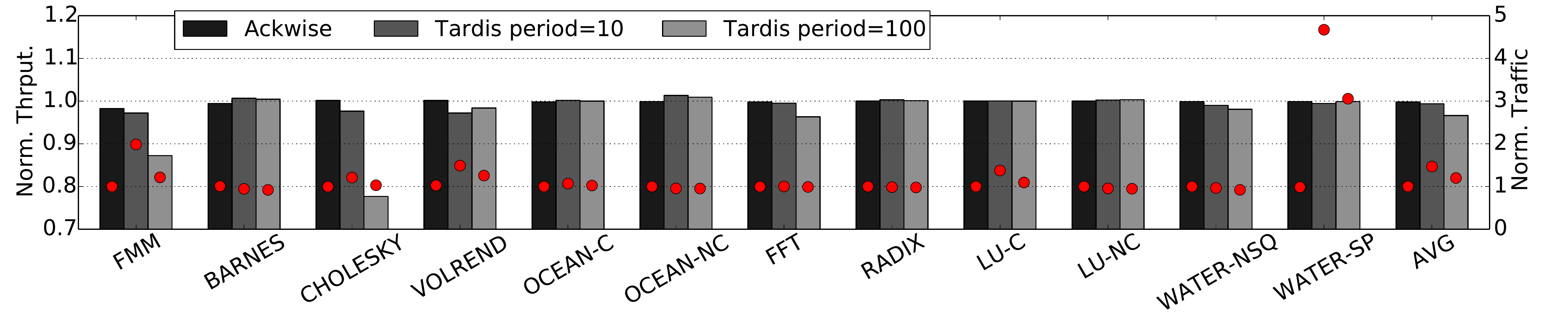}
	} \vspace{-.1in}
	\caption{ Performance of Tardis on 16 and 256 cores.}
	\vspace{-.1in}
	\label{fig:scalability}
\end{figure}

\cref{fig:scalability} shows the performance of Tardis on 16 and 256 
cores respectively. 

At 16 cores, the same configurations are used as at 64 cores. On 
average, the throughput is within 0.2\% of baseline MSI and the 
network traffic is 22.4\% more than the baseline MSI.

At 256 cores, 
two Tardis configurations are shown with self increment period $10$ 
and $100$.
For most benchmarks, both Tardis configurations achieve similar 
performance.  For {\tt FMM}, {\tt CHOLESKY}, however, performance is 
worse when the period is set to 100. As discussed in 
\cref{sec:eval-period}, both benchmarks heavily rely on spinning for 
synchronization. At 256 cores, spinning becomes the system bottleneck 
and period $=100$ significantly delays the spinning core from 
observing the updated variable. It is generally considered bad 
practice to heavily use spinning at high core count.

On average, Tardis with period $=100$ performs 3.4\%  worse than MSI 
with 19.9\% more network traffic. Tardis with period $=10$ makes the
performance 0.6\% within baseline MSI with 46.7\% traffic overhead.

\begin{table}
	\caption{ Storage overhead of different coherence protocols (bits 
	per LLC cacheline) with 4 sharers for Ackwise at 16/64 and 8 sharers at 256 cores.}
    \begin{center}
	{ \scriptsize
		\begin{tabular}{|c|c|c|c|}
            \hline
			\# cores ($N$) & full-map MSI & Ackwise & Tardis \\
			\hline\hline
			16 & 16 bits & 16 bits & 40 bits \\ \hline
			64 & 64 bits & 24 bits & 40 bits \\ \hline
			256 & 256 bits & 64 bits & 40 bits \\ \hline
		\end{tabular}
    }
    \end{center}
	\label{tab:scalability}
    \vspace{-0.25in}
\end{table}

Scalable storage is one advantage of Tardis over directory 
protocols. \cref{tab:scalability} shows the per cacheline storage 
overhead in the LLC for two directory baselines and Tardis. Full-map 
MSI requires one bit for each core in the system, which is $O(N)$ bits 
per cacheline. Both Ackwise and Tardis can achieve $O(\log{N})$ 
storage but Ackwise requires broadcasting support and is thus more 
complicated to implement.

Different from directory protocols, Tardis also requires timestamp 
storage for each L1 cacheline. But the per cacheline storage overhead 
does not increase with the number of cores. 

\subsubsection{Timestamp Size} \label{sec:tssize}

\begin{figure}[t]
	\centering
	\includegraphics[width=\columnwidth]{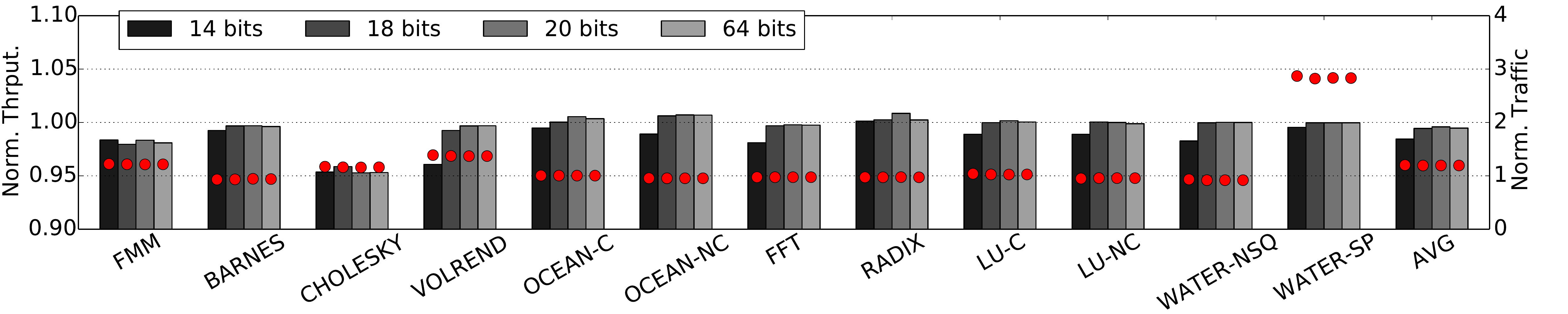}
	\vspace{-.3in}
	\caption{ Performance of Tardis with different timestamp size.}
	\vspace{-.1in}
	\label{fig:tssize}
\end{figure}

\cref{fig:tssize} shows Tardis's performance with different timestamp 
sizes. All numbers are normalized to the baseline MSI. As discussed in 
\cref{sec:opt-compress}, short timestamps roll over more frequently,
which degrades performance due to the rebase overhead. According to 
the results, at 64 cores, 20-bit timestamps can achieve almost the 
same performance as 64-bit timestamps (which never roll over in 
practice). 

\subsubsection{Lease}

\begin{figure}[t]
	\centering
	\includegraphics[width=\columnwidth]{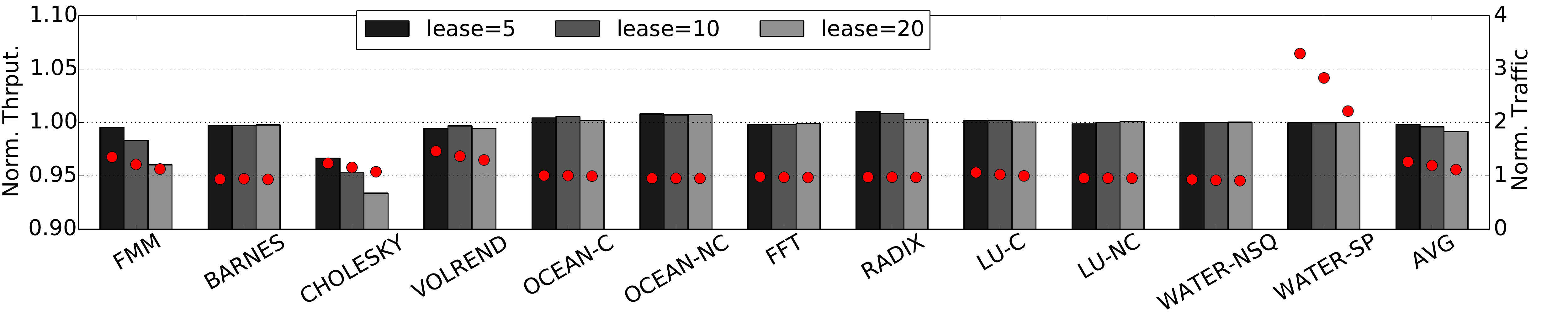}
	\vspace{-.3in}
	\caption{ Performance of Tardis with different lease.}
	\vspace{-.2in}
	\label{fig:lease}
\end{figure}

Finally, we sweep the lease in \cref{fig:lease}. Similar to the self 
increment period, the lease controls when a cacheline expires in the L1 
cache. Roughly speaking, a large lease is equivalent to a long self increment 
period. For benchmarks using a lot of spinning, performance degrades 
since an update is deferred longer. The network traffic also goes down 
as the lease increases. For most benchmarks, however, performance is not 
sensitive to the choice of lease. However, we believe that intelligently choosing
leases can appreciably improve performance; for example, data that is read-only
can be given an infinite lease and will never require renewal. We defer the
exploration of intelligent leasing to future work.

\section{Related Work} \label{sec:related}

We discuss related works on timestamp based coherence protocols
(\cref{sec:related-tscc}) and scalable directory coherence protocols
(\cref{sec:related-scale-dircc}). 

\subsection{Timestamp based coherence} \label{sec:related-tscc}

To the best of our knowledge, none of the existing timestamp based 
coherence protocols is as simple as Tardis and achieves the same level 
of performance as Tardis. In all of these protocols, the timestamp 
notion is either tightly coupled with physical time, or these 
protocols rely on broadcast or snooping for invalidation.

Using timestamps for coherence has been explored in both 
software~\cite{min1989} and hardware~\cite{nandy1994}.
TSO-CC~\cite{elver2014} proposed a hardware coherence protocol based 
on timestamps. However, it only works for the TSO consistency model, 
requires broadcasting support and frequently self-invalidates data in 
private caches. The protocol is also more complex than Tardis. 

In the literature we studied, Library Cache Coherence (LCC) 
\cite{lis2011} is the closest algorithm to Tardis. Different from 
Tardis, LCC uses the physical time as timestamps and requires
a globally synchronized clock. LCC has bad performance because a write 
to a shared variable in LCC needs to wait for all the shared copies to 
expire which may take a long time.  This is much more expensive than 
Tardis which only updates a counter without any waiting.
Singh et al. used a variant of LCC on GPUs with performance 
optimizations~\cite{singh2013}.  However, the algorithm only works 
efficiently for release consistency and not sequential consistency.  

Timestamps have also been used for verifying directory coherence 
protocols~\cite{plakal1998}, for ordering network messages in a 
snoopy coherence protocol~\cite{martin2000}, and to build write-through
coherence protocols \cite{bisiani1989,williams:2000}.
None of these works built coherence protocols purely based on timestamps.
Similar to our work, Martin et. al \cite{martin2000} give a scheme where
processor and memory nodes process coherence
transactions in the same logical order, but not necessarily in the same physical time order.
The network assigns each transaction
a logical timestamp and then broadcasts it to all processor and memory nodes without
regard for order, and the network is required to meet logical time deadlines.
Tardis requires neither broadcast nor network guarantees.
The protocol of \cite{bisiani1989} requires maintaining absolute time across the different processors, and the protocol of \cite{williams:2000} assumes isotach networks \cite{isotach1997}, where all messages travel the same logical distance in the same logical time.

\subsection{Scalable directory coherence} 
\label{sec:related-scale-dircc}

Some previous works have proposed techniques to make directory 
coherence more scalable. Limited directory schemes (\eg, 
\cite{agarwal1998}) only track a small number of sharers and rely on 
broadcasting~\cite{ATAC} or invalidations when the number of sharers 
exceeds a threshold. Although only $O(\log{N})$ storage is required 
per cacheline, these schemes incur performance overhead and/or require
broadcasting which is not a scalable mechanism. 

Other schemes have proposed to store the sharer information in
a chain~\cite{chaiken1990} or hierarchical structures~\cite{maa1991}.
Hierarchical directories reduce the storage overhead by storing the 
sharer information as a $k$-level structure with $\log_k{N}$ bits at 
each level. The protocol needs to access multiple places for each 
directory access and thus is more complex and harder to verify.  

Previous works have also proposed the use of coarse 
vectors~\cite{gupta1990}, sparse directory~\cite{gupta1990}, software 
support~\cite{chaiken1991} or disciplined programs~\cite{choi2011} for 
scalable coherence. Recently, some cache coherence protocols have been 
proposed for 1000-core processors~\cite{kelm2010, sanchez2012}.
These schemes are directory based and require complex 
hardware/software support. In contrast, Tardis can achieve similar 
performance with a very simple protocol.

\section{Conclusion} \label{sec:conclusion}

We proposed a new memory coherence protocol, Tardis, in this paper.  
Tardis is directly derived from the sequential consistency model.  
Compared to popular directory coherence protocols, Tardis is 
simpler to implement and validate, and has better scalability.
Tardis matches the baseline directory protocol in performance in the benchmarks we
evaluated.  For these reasons, we believe Tardis to be a competitive coherence protocol for 
future massive-core and large-scale shared memory systems.

{
	\bibliographystyle{IEEEtran}
	\bibliography{paper}
}

\end{document}